\documentclass[a4paper,11pt]{article}

\usepackage{pos}
\usepackage{caption}
\usepackage{subcaption}
\usepackage{siunitx}
\usepackage{wrapfig}
\usepackage{upgreek}
\usepackage{marvosym}
\usepackage{multirow} 
\usepackage{setspace} 

\newcommand{\fig}[1]{\hyperref[#1]{figure \ref*{#1}}}
\newcommand{\tab}[1]{\hyperref[#1]{table \ref*{#1}}}
\newcommand{\dd}{\ensuremath{\mathrm{d}}}

\newcommand{\spl}{\textit{SplineMPE}}
\newcommand{\ssp}{\textit{SegmentedSplineReco}}
\newcommand{\ssps}{\textit{SegmentedSpline}}
\newcommand{\rnn}{\textit{CRNN-Reco}}

\title{Testing the Pointing of IceCube Using the Moon Shadow in Cosmic-Ray-Induced Muons}
\ShortTitle{Cosmic-Ray Moon Shadow}

\author{The IceCube Collaboration \\{\normalsize \normalfont(a complete list of authors can be found at the end of the proceedings)}}

\emailAdd{saskia.philippen@icecube.wisc.edu}
\emailAdd{thorsten.gluesenkamp@fau.de}
\emailAdd{sebastian.schindler@fau.de}

\abstract{

The IceCube Neutrino Observatory is a cubic-kilometer-scaled detector located at the Geographic South Pole. The calibration of the directional reconstruction of neutrino-induced muons and the pointing accuracy of the detector have to be verified. For these purposes, the moon is used as a standard candle to not rely exclusively on simulated data: Cosmic rays get absorbed by the moon, which leads to a deficit of cosmic-ray-induced muons from the lunar direction that is measured with high statistics. The moon shadow analysis uses an unbinned maximum-likelihood method, which has been methodically improved, and uses a larger detector compared to previous analyses. This allows to observe the shadow with a large significance per month. In the first part, it is found that incorporating a moon disk model, a coordinate-dependent uncertainty scaling and an improved background estimation increase the significance compared to a previous more simplistic analysis. In the second part, the performance of two new directional muon reconstruction algorithms is verified.\\

\vspace{4mm}
{\bfseries Corresponding authors:}
Saskia Philippen$^{1*}$, Thorsten Glüsenkamp$^{2}$, Sebastian Schindler$^{2}$\\
{$^{1}$ \itshape III. Physikalisches Institut B, RWTH Aachen University, D-52074 Aachen, Germany}\\
{$^{2}$ \itshape Erlangen Centre for Astroparticle Physics, Friedrich-Alexander-Universit{\"a}t Erlangen-N{\"u}rnberg, D-91058 Erlangen, Germany}\\[4mm]
$^*$ Presenter

}

\FullConference{37$^{\rm{th}}$ International Cosmic Ray Conference (ICRC 2021)\\
		July 12th -- 23rd, 2021\\
		Online -- Berlin, Germany}

\begin{document}
\maketitle

\section{Introduction}

IceCube is a cubic-kilometer neutrino detector located at the Geographic South Pole, which consists of over 5000 light detection units placed more than one kilometer deep into the clear ice \cite{DetectorPaper}. Cherenkov photons emitted by charged particles form the detectable signal, whose time and location information allows to reconstruct particle type, energy and direction. Interactions of muon-neutrinos in the medium produce secondary muons, which are detected as tracks resolvable in space and time. This allows for a good reconstruction of their travel directions with an angular resolution of about $\ang{1}$ \cite{7yrPaper}, making muons indispensable for the reconstruction of the neutrino origin. However, the majority of recorded events stem from muons that are created in air showers induced by cosmic rays in the atmosphere, and which penetrate the thick ice shield above the detector if their energy is sufficiently high.

Cosmic rays arrive to first order isotropically at Earth and have high enough energies, such that cosmic-ray-induced muons that are detected by IceCube travel in roughly the same direction. Consequently, the muon flux in the detector is mostly uniform in azimuth, and only dependent on zenith due to the differing amount of material traversed by the muons. However, as the moon blocks any cosmic rays arriving from that direction in the sky, a deficit in the flux can be observed \cite{MoonPaper}. This moon shadow acts as a standard candle for muon pointing, and can be used for several applications. Among them are tests of different analysis techniques without the need to rely on Monte Carlo simulations, or a regular testing of the detector performance. 

In this paper, we first present some improved methodologies of the moon shadow analysis over previous analyses that were shown in \cite{MoonPaper}. In the second part, the accuracy of two new muon track reconstruction algorithms, \ssp{} \cite{SegSplinePaper} and a machine-learning-based reconstruction using a neural network with convolutional and recurrent elements (\rnn{}), are compared to the prevailing algorithm using the improved moon shadow analysis.

\section{Moon Shadow Analysis Methods}
\subsection{Analysis Principle}
The basic principle of the moon shadow analysis is to measure a deficit in cosmic-ray-induced muons. To achieve this, the source hypothesis is tested on a $\pm \ang{3}$ grid moving with the moon by comparing the events in the $\pm \ang{10}$ window around the moon (on-source region) to the true background distribution, which is determined from an off-source region in the same zenith band but at different azimuth values. The analysis is done in a quasi-Cartesian coordinate system defined by the azimuth and zenith angles relative to the moon position $\Delta\phi=(\phi-\phi_\text{\Moon})\cdot\sin(\theta)$ and $\Delta\theta=\theta-\theta_\text{\Moon}$, respectively. The significance to measure the moon at a grid point is determined by using a maximum-likelihood method, with the likelihood function defined as 
\begin{align}
&\mathrm{log}\mathcal{L}\left(n_\mathrm{s}, \Delta\phi, \Delta\theta|\vec{x}_{1..N}, \boldsymbol{\Sigma}'_{1..N}\right) = \sum_{i=1}^{N} \log\left(
\frac{n_\mathrm{s}}{N}\widetilde{S}(\Delta\phi, \Delta\theta|\vec{x}_i, \boldsymbol{\Sigma}'_i)+
\left(1-\frac{n_\mathrm{s}}{N} \right) \widetilde{B}(\vec{x}_i, \boldsymbol{\Sigma}'_i)
\right),
\end{align}
where $\boldsymbol{\Sigma}'_i$ is the covariance matrix and $\vec{x_i} = \begin{pmatrix}\delta\phi_i\\\delta\theta_i\\\end{pmatrix} $ is the positional vector from the reconstructed direction to the grid point for the $i$\textsuperscript{th} event. Using this function, the number of events caused or blocked by a source $n_\mathrm{s}$ is fit with regards to the total number of events in the source region $N$, where the source term $\widetilde{S}(\vec{x}_i, \boldsymbol{\Sigma}'_i)$ and background term $\widetilde{B}(\vec{x}_i, \boldsymbol{\Sigma'}_i)$ are explained in section \ref{sec:signal} and \ref{sec:bgd}, respectively.

\subsubsection{Event Uncertainty Estimation}
The uncertainties of the directional reconstructions of the muons are often approximated by two-dimensional asymmetric Gaussian distributions in the likelihood landscape \cite{paraboloid}. The contours are ellipses described with the semi-major and -minor axes $\sigma_1,\ \sigma_2$ and the rotational angle $\alpha$ counterclockwise to the azimuth axis. The uncertainties are typically underestimated, and an energy-dependent global re-scaling using CORSIKA simulations \cite{corsika} is performed to obtain better statistical coverage. In contrast to the standard methodology with a symmetricized uncertainty approximation using global re-scaling, this section introduces a new re-scaling scheme motivated by the detector geometry which works for general ellipse orientations. The scaling is done separately in the azimuth and zenith axes with the factors $s_\phi$ and $s_\theta$, leading to the scaled two-dimensional Gaussian distribution centered at the origin of the coordinate system, which equals the reconstructed event direction,
\begin{alignat}{2}
f_\mathrm{2D}(\vec{x}, \boldsymbol{\Sigma}, s_\phi, s_\theta)&= \frac{\mathrm{e}^{-\frac{a\cdot \delta\phi^2 + b\cdot \delta\theta^2 + c\cdot \delta\phi\delta\theta}{2(s_\phi s_\theta\sigma_1\sigma_2)^2} }}{2\pi s_\phi s_\theta\sigma_1\sigma_2}\ \mathrm{with}\quad a(\boldsymbol{\Sigma}, s_\theta) = s_\theta^2\left(\sigma_1^2\sin(\alpha)^2 + \sigma_2^2  \cos(\alpha)^2  \right),
\label{eqGaussian2D}\\
b(\boldsymbol{\Sigma}, s_\phi) &= s_\phi^2 \left( \sigma_1^2 \cos(\alpha)^2  +  \sigma_2^2\sin(\alpha)^2 \right)\quad\mathrm{and}\quad
c(\boldsymbol{\Sigma},s_\phi, s_\theta) = s_\phi s_\theta\sin(2\alpha)\left(\sigma_2^2 - \sigma_1^2 \right),\nonumber
\end{alignat}
using the covariance matrix $\mathbf{\Sigma'} = \mathbf{S}\mathbf{\Sigma} \mathbf{S} = \mathbf{S}\mathbf{R} \mathbf{\Lambda} \mathbf{R^T} \mathbf{S}$ in a scaled eigenvalue decomposition, with
\begin{align*}
    \text{scaling matrix }
    \mathbf{S} = \begin{pmatrix} 
        s_\phi & 0 \\
        0 & s_\theta\\
    \end{pmatrix}, \quad
    \mathbf{R} = \begin{pmatrix}
        \cos(\alpha) & -\sin(\alpha)\\
        \sin(\alpha) & \cos(\alpha)\\
    \end{pmatrix}, \quad
    \mathbf{\Lambda} = \begin{pmatrix}
        \sigma_1^2 & 0\\
        0 & \sigma_2^2\\
    \end{pmatrix}.
\end{align*}
The 2D-Gaussian distribution is projected onto the main axes by marginalization, resulting in 1D-Gaussian distributions with standard deviations 
\begin{align}
\sigma_\phi(\boldsymbol{\Sigma}, s_\phi)&=\sqrt{b}
\quad \mathrm{and}\quad \sigma_\theta(\boldsymbol{\Sigma},s_\theta)=\sqrt{a}
\end{align}
Using the 1D-distributions, $s_\phi$ and $s_\theta$ are determined such that for all events in a certain energy range, 68.27\% of these events have an angular difference between their true and reconstructed directions smaller or equal than their individual values for $\sigma_\phi$ and $\sigma_\theta$ when scaled by $s_\phi$ and $s_\theta$, respectively. To estimate the quality of this scaling, two criteria are verified in the two-dimensional distribution:\\
(1) for 39.34\% of all events the true direction should lie in the $1\sigma$ contour.\\
(2) for 50\% of all events the true direction should lie in the $\sqrt{\log(2)}\sigma = 1.177\sigma$ contour. \\
The ellipse equation of a contour of the Gaussian after scaling is calculated as follows:\\
The stretched covariance matrix $\mathbf{\Sigma'}=\mathbf{S}\mathbf{R} \mathbf{\Lambda} \mathbf{R}^T \mathbf{S}$ can be expressed as $\boldsymbol{\mathcal{R}} \mathbf{\Lambda'} \boldsymbol{\mathcal{R}}^{T}$, where $\boldsymbol{\mathcal{R}}$ is the rotation matrix for the angle $\alpha'$ as defined below, and the diagonal matrix $\mathbf{\Lambda'}$ gives the semi-major and -minor axes $\sigma_1'$ and $\sigma_2'$ of the scaled ellipse via its eigenvalues.
\begin{align}
\sigma_{1/2}' &= \frac{a+b\pm d}{2}\quad\mathrm{and}\quad \alpha' = \left\{
\begin{array}{ll}
\beta + \frac{\pi}{2} & \alpha < \pi \\
\beta + \frac{3\pi}{2} & \alpha > \pi \\
\end{array}\right.\nonumber\\
&\mathrm{with}\quad d=\sqrt{(a+b)^2-4(\sigma_1\sigma_2s_\phi s_\theta)^2}\quad \mathrm{and}\quad\beta = -\frac{1}{\sqrt{1+\frac{(a - b - d)^2}{c^2}}} \nonumber\\
\Rightarrow& \quad \sigma_\mathrm{2D}(\Delta\varphi,\Delta\vartheta)=\frac{\left(\Delta\varphi\cdot\cos\left(\alpha'\right)+\Delta\vartheta\cdot\sin\left(\alpha'\right)\right)^2}{\sigma_1'^2}+\frac{\left(-\Delta\varphi\cdot\sin\left(\alpha'\right)+\Delta\vartheta\cdot\cos\left(\alpha'\right)\right)^2}{\sigma_2'^2}
\end{align}
This describes the statistical coverage in two dimensions, where $\Delta\varphi$ and $\Delta\vartheta$ are the differences between reconstructed and true directions in the Cartesian coordinate system.

\begin{figure}
	\includegraphics[width=1.\textwidth]{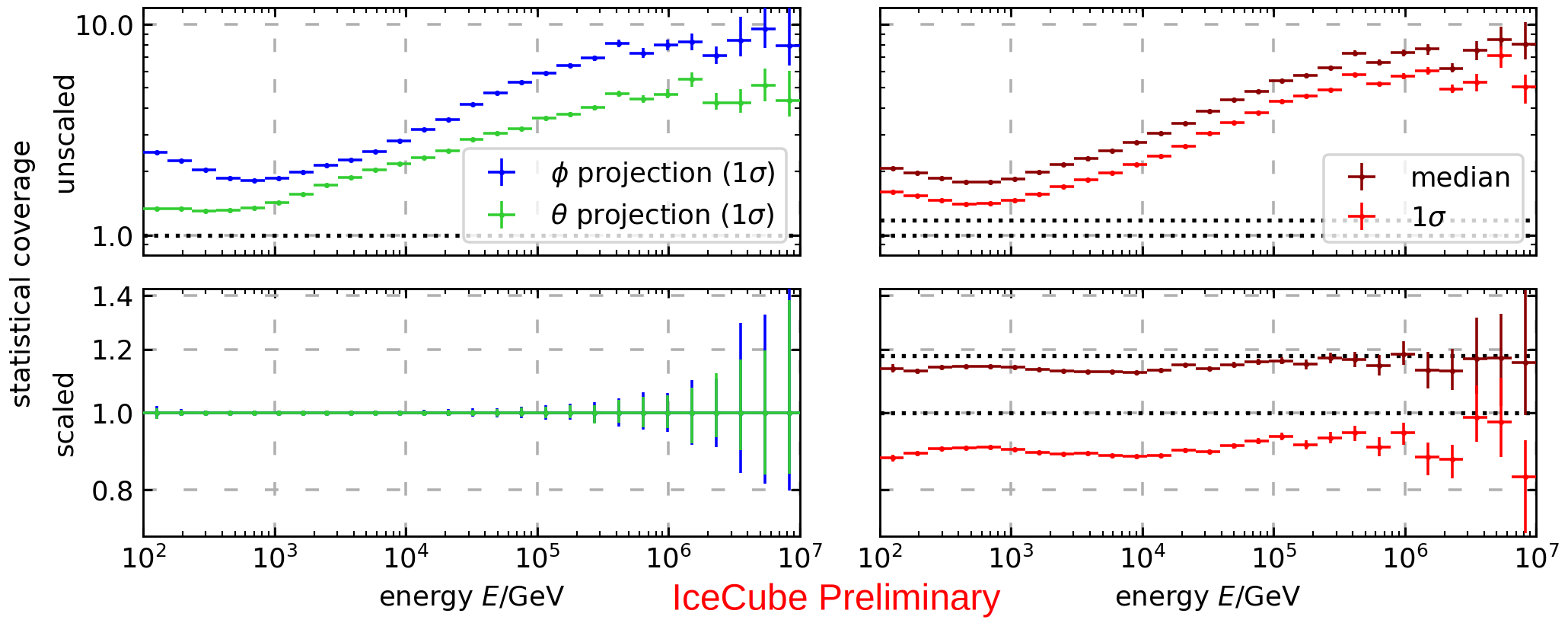}
	\caption{Actual statistical coverage in units of the nominal statistical coverage $\sigma$, therefore the $1\sigma$ line should lie at one. Left plots show the projections onto the main axes, right ones show the 2D case. The upper plots show the values for the unscaled, the lower ones for the scaled uncertainties.}
	\label{figPullEnergy}
\end{figure}

As shown in \fig{figPullEnergy}, the separate scaling in azimuth and zenith directions fulfills the criteria for the projections per definition, and both test criteria in two dimensions fit very well with uncertainties between $0.88\sigma$ and $0.98\sigma$ for the $1\sigma$ contour and between $1.13\sigma$ and $1.18\sigma$ for the median.

\subsubsection{Background Term} \label{sec:bgd}
The azimuth dependence of the event rate is mostly constant due to the movement of the moon in local coordinates. Thus, an off-source region in azimuth allows to determine the background distribution, which is needed to counteract the strong dependence in zenith direction.
The true background distribution is determined as the sum of the scaled Gaussian uncertainty distributions of all events evaluated on grid points relative to the center position of the off-source region, which is then normalized. In the used likelihood formation, edge effects are unavoidable, as long as uncertainty ellipses around the moon region stretch significantly out of the actual background window. Therefore, events with large semi-major axes $\sigma_1'$ are discarded, and the normalization per radiant squared has to be performed  with a safety margin (\ang{3} for a $\sigma_1' < \ang{1}$ cut for this work).

Finally, instead of evaluating the previously determined background distribution, the probability for an event in the source region to be background is calculated as the expected value of the background distribution under the event's Gaussian distribution:
\begin{align}
\widetilde{B}(\vec{x}_i, \boldsymbol{\Sigma'}_i) = \mathrm{E}[B(x)]_{f_{2D}(x|\vec{x}_i, \boldsymbol{\Sigma'}_i)}.
\end{align}

\subsubsection{Source Term} \label{sec:signal}
For analyzing point-like sources, the source term of an event is typically given by evaluating its Gaussian uncertainty distribution at a grid point, which is equivalent to integration with a delta distribution. Following this same ansatz, the source hypothesis for an extended source, which appears as a disc with radius $R_\text{\Moon}$ in the two-dimensional projection, can be described by integration with a Heaviside step function (for calculation and solution see \cite{SaskiasMoon}):
\begin{align}
\widetilde{S}(\vec{x}, \boldsymbol{\Sigma'}) &=\frac{1}{\pi\frac{
		R_\text{\Moon}^2}{\sin(\theta)}}\int_{-\infty}^{\infty}\dd\widehat{\delta
\phi}\int_{-\infty}^{\infty}\dd\widehat{\delta \theta}\  \Uptheta\left( \begin{pmatrix}|\delta\phi-\widehat{\delta\phi}|\\|\delta\theta-\widehat{\delta\theta}|\\\end{pmatrix}  \leq  \begin{pmatrix}R_\text{\Moon}\sin(\theta)^{-1}\\(R_\text{\Moon}^2 -
		(\delta\phi-\widehat{\delta\phi})^2\sin(\theta)^2)^{\frac{1}{2}}\\\end{pmatrix}   \right) 
\cdot f_\mathrm{2D}(\widehat{\vec{x}}, \boldsymbol{\Sigma'})
\end{align}
where the result is numerically solvable. The radius in azimuth direction is scaled with $\frac{1}{\sin(\theta)}$ to transform the circular moon into the quasi-Cartesian coordinate system.

\subsection{Evaluation}
The result of the maximum likelihood method is a landscape of the best fit values of $n_\mathrm{s}$ at every grid point. To interpret the result, two quantities are introduced using Wilks' theorem \cite{Wilks}:\\
(1) The \textit{source significance} describes the deviation from the background hypothesis, indicating how well a source can be determined. This quantity is calculated with the logarithmic likelihood difference with respect to $n_\mathrm{s}=0$ for one degree of freedom, and is given the opposite sign of $n_\mathrm{s}$. This way, a sink like the moon has a positive significance, which will be called a source in the following for simplicity. Because relative comparisons are performed, a trial factor is not calculated.\\
(2) The \textit{pointing significance} indicates the precision of the positional reconstruction. It is determined with the logarithmic likelihood difference with respect to the global minimum in $n_\mathrm{s}$ with two degrees of freedom due to its position. Together with the known position of the moon, it shows the pointing accuracy, which however is impacted by the geomagnetic field.

\subsection{Performance Test}
The analysis presented in this paper is an improved version of a previous analysis \cite{MoonPaper}. The maximum source significance improves from $12.17\sigma$ to $13.46\sigma$ when applied to the same eight-months data sample using the prevailing muon reconstruction algorithm. Removing each of the individual improvements, while keeping the others, yields reduced source significances: $13.08\sigma$ for the uncertainties estimation and $13.02\sigma$ for the source model.

\section{Test of Directional Reconstruction Algorithms}
\subsection{Analysis}
In the following, we use two different methods to compare the precision of two different muon reconstructions. The first method uses reconstruction-dependent cuts on the respective uncertainty estimate, which leads to a different number of events per reconstruction. We call this the \emph{individual} method. In order to have the same number of events for a direct comparison, the second method involves taking the subset of the events that pass both reconstruction-dependent cuts, and is called the \emph{intersection} method. While it offers a more direct comparison, it has the disadvantage that a reconstruction with small uncertainties will be punished due to the larger uncertainties of the reconstruction it is being compared to.

The used data set comprises the first five moon cycles in 2013, which are each about \num{12} successive days during which the moon is visible at the South Pole. The baseline for the comparison of two new reconstruction algorithms is \textbf{\spl{}} \cite{SplineMPE}, which is the current default used for muon track reconstruction in IceCube. It assumes a continuous muon energy loss and can handle stochastic losses only effectively, which is a simplification addressed by the recently developed \textbf{\ssp{}} \cite{SegSplinePaper}. This newer reconstruction models the stochastic energy losses of high-energy muons explicitly. The third reconstruction is the \textbf{\rnn{}} \cite{RNNInternal}, which is currently in development. It uses a neural network trained on Monte Carlo simulations to directly predict the muon direction and its uncertainty.

\rnn{} provides only symmetrical uncertainty estimates. It is compared to \spl{} with symmetricized Gaussian uncertainties using a symmetrical uncertainty re-scaling, while \ssp{} is independently compared to \spl{} with asymmetric uncertainties and geometry-motivated re-scaling. The algorithms can be compared to each other on the five moon cycles individually, and on the combined five-cycle data set for a statistically more robust result.

\begin{table}
	\centering
	\small
	\begin{tabular}{l|l|l||l|l|l|l|l||l}
		cut method                      & uncertainty                   & reconstruction    & 1				& 2				& 3				& 4				& 5				& combined	        \\ \hline \hline
		\multirow{4}{*}{intersection}   & \multirow{2}{*}{asymmetric}	& \spl{}            & $6.0 \sigma$	& $3.0 \sigma$	& $3.8 \sigma$	& $6.0 \sigma$	& $5.5 \sigma$	& $10.2 \sigma$     \\
            		                    & 		                        & \ssps{}           & $5.1 \sigma$	& $3.0 \sigma$	& $3.7 \sigma$	& $6.4 \sigma$	& $5.5 \sigma$	& $10.0 \sigma$     \\ \cline{2-9}
		                                & \multirow{2}{*}{symmetric}	& \spl{}            & $7.3 \sigma$	& $5.2 \sigma$	& $5.0 \sigma$	& $7.1 \sigma$	& $6.7 \sigma$	& $14.3 \sigma$     \\
            		                    & 	            	            & \rnn{}            & $6.3 \sigma$	& $4.4 \sigma$	& $4.7 \sigma$	& $5.4 \sigma$	& $6.6 \sigma$	& $12.7 \sigma$     \\ \hline \hline
		\multirow{4}{*}{individual}     & \multirow{2}{*}{asymmetric}   & \spl{}            & $5.7 \sigma$	& $3.6 \sigma$	& $4.5 \sigma$	& $6.6 \sigma$	& $5.6 \sigma$	& $11.2 \sigma$     \\
            		                    & 		                        & \ssps{}           & $5.3 \sigma$	& $3.5 \sigma$	& $3.7 \sigma$	& $6.5 \sigma$	& $5.0 \sigma$	& $10.0 \sigma$ 	\\ \cline{2-9}
		                                & \multirow{2}{*}{symmetric} 	& \spl{}            & $7.4 \sigma$	& $5.4 \sigma$	& $5.1 \sigma$	& $7.9 \sigma$	& $6.9 \sigma$	& $15.1 \sigma$     \\
            		                    & 		                        & \rnn{}            & $6.5 \sigma$	& $4.4 \sigma$	& $5.4 \sigma$	& $5.7 \sigma$	& $7.3 \sigma$	& $13.7 \sigma$     \\
	\end{tabular}
	\caption{\textcolor{red}{IceCube Preliminary}. Source significances for five moon cycles and the combined data set. The top half shows both comparisons with the \emph{intersection} cut method, the lower half all four reconstructions using \emph{individual} cuts. Cutting on different uncertainties also affects the background distribution, which can lead to larger significances in the \textit{intersection} method, even though less events are analyzed. Note that asymmetric and symmetric \spl{} cannot be compared due to the different amount of events.}
	\label{tabAlgosResults}
\end{table}

\subsection{Discussion}
\begin{figure}
	\centering
	\includegraphics[scale=.19]{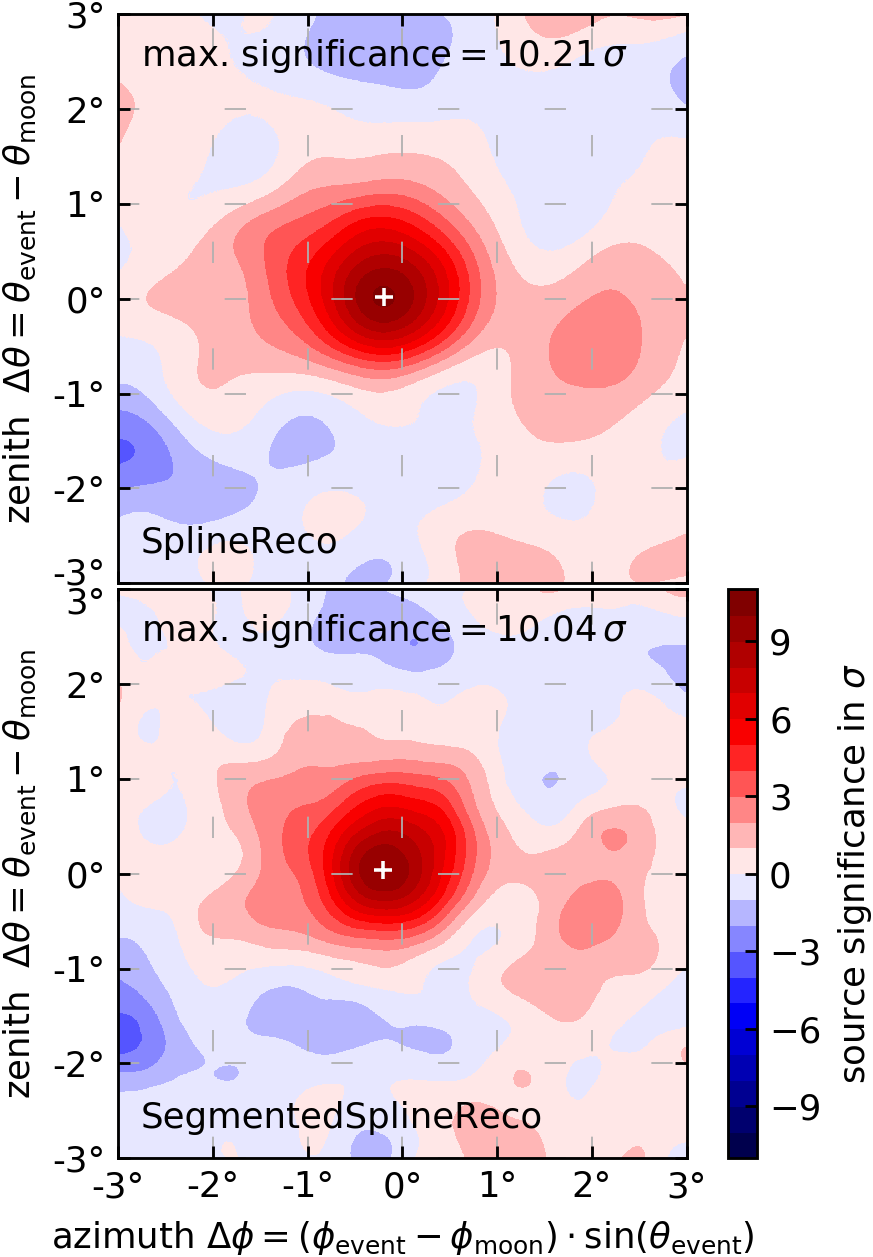}
	\hfill \vrule \hfill
	\includegraphics[scale=.19]{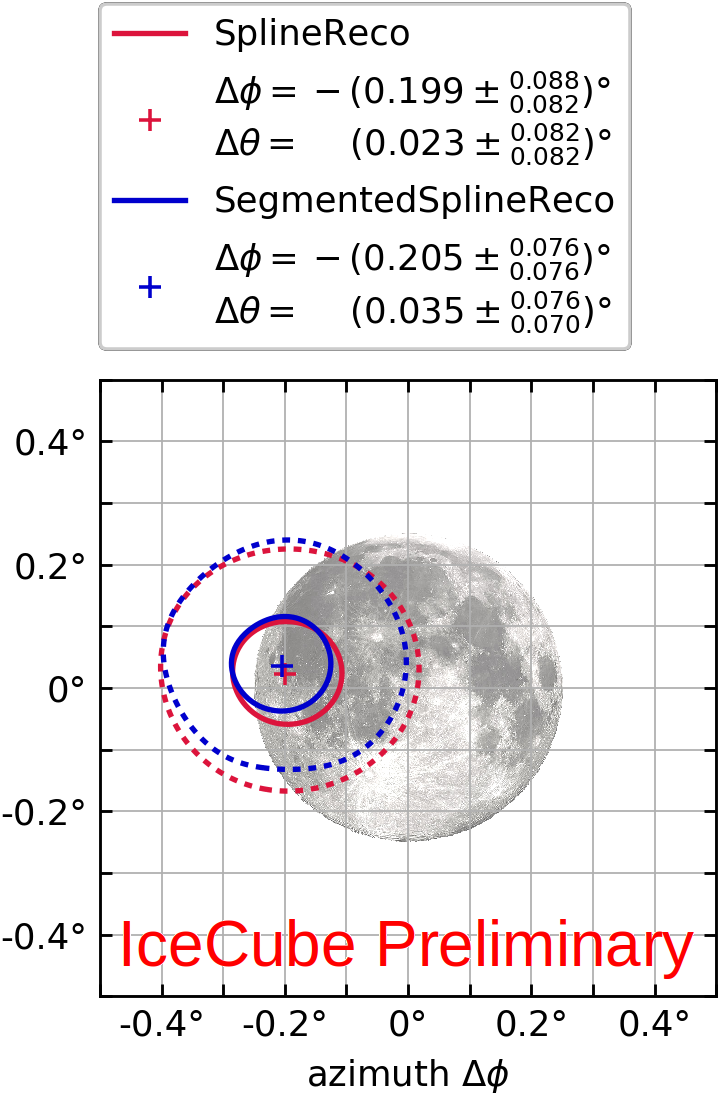}
	\hfill
	\includegraphics[scale=.19]{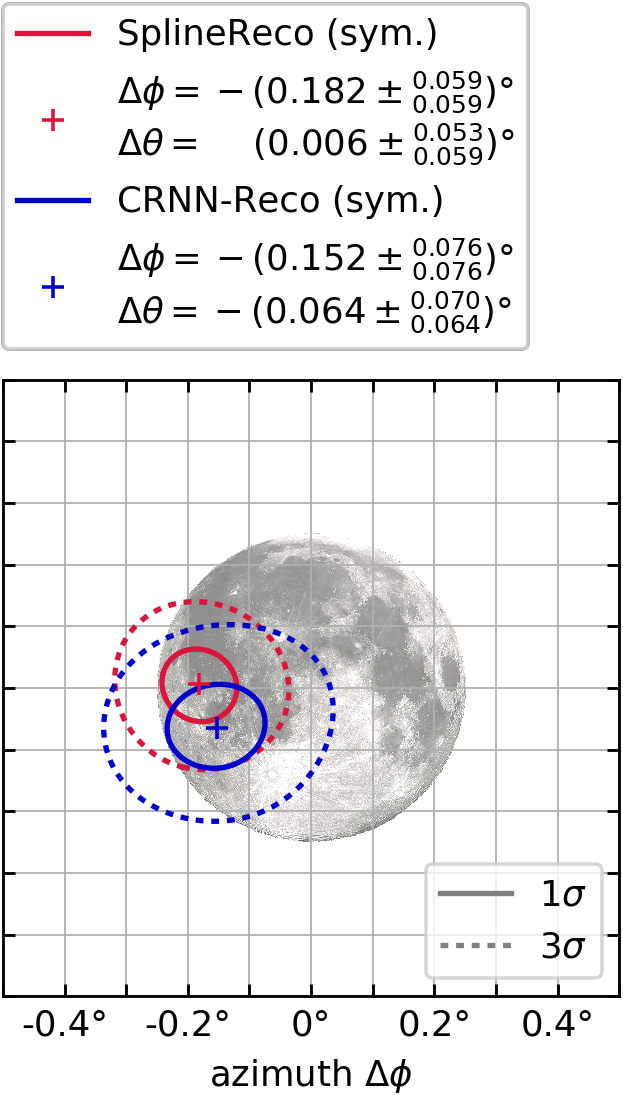}
	\caption{Source significance landscapes (left) shown exemplary only for \spl{} compared to \ssp{}, and contours of the pointing significance for both \ssp{} (middle) and \rnn{} (right) comparisons. Crosses mark the points of maximum likelihood $(\Delta \phi, \Delta \theta)$. Shown is the combined data set with \emph{intersection} cuts.}
	\label{figRecoCombined}
\end{figure}

The moon shadow is observed by all reconstruction algorithms in all five cycles with varying source significances, displayed in \tab{tabAlgosResults}. For a better understanding, \fig{figRecoCombined} shows the source significance landscape with a clearly visible moon shadow and surrounding background fluctuations. These are similarly distributed between the compared reconstructions, which indicates that they come from the data itself and not from reconstruction differences. The pointing significance fluctuates strongly between reconstructions on single moon cycles, but is a reliable result on the combined data set as shown in \fig{figRecoCombined}. All reconstructions are compatible within their $1 \sigma$ contours, and show a systematic shift of up to \ang{0.2} to smaller azimuth values, which might be attributed to the geomagnetic field or other systematic effects.

Using the \emph{intersection} method, the source significance for \rnn{} is smaller than for \spl{} for all single cycles and for the combined data set. When using \emph{individual} cuts, the picture does not change much, which supports this result. The precision of the reconstruction, as indicated by the size of the pointing significance contours, is smaller than \spl{}. Overall, this indicates that \rnn{} is performing worse than \spl{} on cosmic-ray-induced muon events. As these consist of bundles of several muons, instead of the single muons \rnn{} is intended to reconstruct and which it was trained on, the performance is likely worse than it potentially could be. Therefore, this result cannot be generalized to muon neutrinos. However, it proves that a machine-learning-based direction reconstruction works well when subjected to real data instead of Monte Carlo simulations.

The differences in source significance between \ssp{} and \spl{} are smaller and without a clear tendency across moon cycles as in the previous comparison, and there is no large difference on the combined data set. \emph{Individual} cuts again do not substantially differ from this observation. The size and position of the pointing contours are very similar to \spl{}. This comparable performance is expected, as the largest advantage for \ssp{} comes from high-energy ($ > \SI{50}{TeV} $) muons, and the moon data set is limited to muon energies mainly between \SIrange{1}{10}{TeV}.

The comparison is still not entirely fair, as in the above result \ssp{} only used a simple uncertainty estimation based on the Hesse matrix that does not include a final time convolution, called \emph{Post-jitter} in \cite{SegSplinePaper}. In the current implementation the uncertainty determination with \emph{Post-jitter} can only be calculated by performing a time consuming Markov Chain Monte Carlo sampling and a subsequent covariance determination from the samples, which was too time consuming for this study.

\section{Summary}
In this paper, first we discussed some extensions to the moon analysis tools in IceCube. We motivated and derived a 2D-based rescaling of the uncertainty contours based on local detector coordinates, which demonstratively leads to better coverage. A new method for the background estimation using an event's uncertainty information, as well as a disk source model were introduced. The performance improvements of these methods are quantified in significance comparisons to the standard method.

In the second half, we tested two new reconstructions. The first is \ssp{}, which we compare to the standard reconstruction \spl{} using the new rescaling technique introduced in the first section. It performs comparably to the existing standard reconstruction in the energy range of the data, which is expected. The second new reconstruction is a machine-learning-based reconstruction, \rnn{}, which is compared to \spl{} using symmetricized uncertainties. It is slightly worse than the likelihood-based standard, but matches expectations from Monte Carlo and does not have a larger systematic bias.

In the future, the moon analysis is planned to be automated for a monthly detector test, and used for further studies, like investigating the geomagnetic field or for tests of calibration improvements.

\bibliographystyle{ICRC}
\bibliography{references}

\clearpage
\section*{Full Authors List: IceCube Collaboration}

\scriptsize
\noindent
R. Abbasi$^{17}$,
M. Ackermann$^{59}$,
J. Adams$^{18}$,
J. A. Aguilar$^{12}$,
M. Ahlers$^{22}$,
M. Ahrens$^{50}$,
C. Alispach$^{28}$,
A. A. Alves Jr.$^{31}$,
N. M. Amin$^{42}$,
R. An$^{14}$
% ...
\\

\noindent
% ...
$^{12}$ Universit{\'e} Libre de Bruxelles, Science Faculty CP230, B-1050 Brussels, Belgium \\
% ...
$^{14}$ Department of Physics and Laboratory for Particle Physics and Cosmology, Harvard University, Cambridge, MA 02138, USA \\
% ...
$^{17}$ Department of Physics, Loyola University Chicago, Chicago, IL 60660, USA \\
$^{18}$ Dept. of Physics and Astronomy, University of Canterbury, Private Bag 4800, Christchurch, New Zealand \\
% ...
$^{22}$ Niels Bohr Institute, University of Copenhagen, DK-2100 Copenhagen, Denmark \\
% ...
$^{28}$ D{\'e}partement de physique nucl{\'e}aire et corpusculaire, Universit{\'e} de Gen{\`e}ve, CH-1211 Gen{\`e}ve, Switzerland \\
% ...
$^{31}$ Karlsruhe Institute of Technology, Institute for Astroparticle Physics, D-76021 Karlsruhe, Germany  \\
% ...
$^{42}$ Bartol Research Institute and Dept. of Physics and Astronomy, University of Delaware, Newark, DE 19716, USA \\
% ...
$^{50}$ Oskar Klein Centre and Dept. of Physics, Stockholm University, SE-10691 Stockholm, Sweden \\
% ...
$^{59}$ DESY, D-15738 Zeuthen, Germany \\
% ...

\subsection*{Acknowledgements}

\noindent
USA {\textendash} U.S. National Science Foundation-Office of Polar Programs,
U.S. National Science Foundation-Physics Division,
U.S. National Science Foundation-EPSCoR,
Wisconsin Alumni Research Foundation,
Center for High Throughput Computing (CHTC) at the University of Wisconsin{\textendash}Madison,
Open Science Grid (OSG),
Extreme Science and Engineering Discovery Environment (XSEDE),
Frontera computing project at the Texas Advanced Computing Center,
U.S. Department of Energy-National Energy Research Scientific Computing Center,
Particle astrophysics research computing center at the University of Maryland,
Institute for Cyber-Enabled Research at Michigan State University,
and Astroparticle physics computational facility at Marquette University;
Belgium {\textendash} Funds for Scientific Research (FRS-FNRS and FWO),
FWO Odysseus and Big Science programmes,
and Belgian Federal Science Policy Office (Belspo);
Germany {\textendash} Bundesministerium f{\"u}r Bildung und Forschung (BMBF),
Deutsche Forschungsgemeinschaft (DFG),
Helmholtz Alliance for Astroparticle Physics (HAP),
Initiative and Networking Fund of the Helmholtz Association,
Deutsches Elektronen Synchrotron (DESY),
and High Performance Computing cluster of the RWTH Aachen;
Sweden {\textendash} Swedish Research Council,
Swedish Polar Research Secretariat,
Swedish National Infrastructure for Computing (SNIC),
and Knut and Alice Wallenberg Foundation;
Australia {\textendash} Australian Research Council;
Canada {\textendash} Natural Sciences and Engineering Research Council of Canada,
Calcul Qu{\'e}bec, Compute Ontario, Canada Foundation for Innovation, WestGrid, and Compute Canada;
Denmark {\textendash} Villum Fonden and Carlsberg Foundation;
New Zealand {\textendash} Marsden Fund;
Japan {\textendash} Japan Society for Promotion of Science (JSPS)
and Institute for Global Prominent Research (IGPR) of Chiba University;
Korea {\textendash} National Research Foundation of Korea (NRF);
Switzerland {\textendash} Swiss National Science Foundation (SNSF);
United Kingdom {\textendash} Department of Physics, University of Oxford.

\end{document}